\documentclass[12pt,a4paper]{article}
\usepackage{color}
\usepackage{fullpage}
\usepackage[centertags]{amsmath}
\usepackage{pstricks}
\allowdisplaybreaks[4]
\usepackage{amssymb}
\usepackage{bm}
\usepackage[dvips]{graphicx}
\usepackage{url}
\usepackage[dvips,hyperindex]{hyperref}
\newcommand{\boss}[2]{\ensuremath{\rlap{\kern-2.5pt\ensuremath{\overset{\scriptscriptstyle(-)}{\phantom{#1}}}}{\ensuremath{{#1}_{#2}}}}}
\begin{document}
\begin{flushright}
\texttt{CERN-PH-TH/2008-238, LAPTH-1294/08}
\\
\textsf{\today}
\end{flushright}
\vspace{1cm}
\begin{center}
\Large\bfseries
Cosmological constraints on a light non-thermal sterile neutrino\\[0.5cm]
\large\normalfont
Mario A. Acero\ensuremath{^{(a,b,c)}}, Julien Lesgourgues\ensuremath{^{(d,e)}}
\\[0.3cm]
\small\itshape
\setlength{\tabcolsep}{0.7pt}
\begin{tabular}{cl}
\ensuremath{(a)}
&
Dipartimento di Fisica Teorica,
Universit\`a di Torino,
\\
&
Via P. Giuria 1, I--10125 Torino, Italy
\\[0.1cm]
\ensuremath{(b)}
&
INFN, Sezione di Torino,
Via P. Giuria 1, I--10125 Torino, Italy
\\[0.1cm]
\ensuremath{(c)}
&
Laboratoire d'Annecy-le-Vieux de Physique Th\'orique LAPTH,
\\
&
Universit\'e de Savoie, CNRS/IN2P3, 74941 Annecy-le-vieux, France
\\[0.1cm]
\ensuremath{(d)}
&
PH-TH, CERN - CH-1211 Geneva 23, Switzerland
\\[0.1cm]
\ensuremath{(e)}
&
SB-ITP-LPPC, BSP 312 - EPFL, CH-1015 Lausanne, Switzerland
\end{tabular}
\end{center}
\begin{abstract}
Although the MiniBooNE experiment has severely restricted the possible
existence of light sterile neutrinos, a few anomalies persist in
oscillation data, and the possibility of extra light species
contributing as a subdominant hot (or warm) component is still
interesting. In many models, this species would be in thermal
equilibrium in the early universe and share the same temperature as
active neutrinos, but this is not necessarily the case.  In this work,
we fit up-to-date cosmological data with an extended $\Lambda$CDM
model, including light relics with a mass typically in the range
0.1--10~eV. We provide, first, some nearly model-independent
constraints on their current density and velocity dispersion, and
second, some constraints on their mass, assuming that they consist
either in early decoupled thermal relics, or in non-resonantly
produced sterile neutrinos.  Our results can be used for constraining
most particle-physics-motivated models with three active neutrinos and
one extra light species. For instance, we find that at the $3\sigma$
confidence level, a sterile neutrino with mass $m_s = 2$ eV can be
accommodated with the data provided that it is thermally distributed
with $T_s/T_\nu^{\rm id} \lesssim 0.8$, or non-resonantly produced
with $\Delta N_{\text{eff}} \lesssim 0.5$. The bounds become
dramatically tighter when the mass increases. For $m_s \lesssim 0.9$ eV
and at the same confidence level, the data is still compatible with a
standard thermalized neutrino.
\end{abstract}

\section{Introduction}
Neutrino oscillation is a well studied phenomenon, confirmed by strong experimental evidences. Most experimental results are well explained with a three-neutrino oscillation model, involving two independent and well-measured square-mass differences: $\Delta m^2_{sol} = (7.59 \pm 0.21) \times 10^{-5}$ eV$^2$ \cite{Abe:2008ee} and $\Delta m_{atm} = (2.74^{+0.44}_{-0.26}) \times 10^{-3}$ eV$^2$ \cite{Adamson:2007gu}. However, some other experiments have shown some anomalies which do not fit in this hypothesis (LSND \cite{Aguilar:2001ty}, Gallium experiments \cite{Abdurashitov:2005tb}, MiniBooNE low energy anomaly \cite{AguilarArevalo:2007it}). These anomalous results might be due to unknown systematic effects, but all attempts to identify such systematics have failed until now. Otherwise, they could be interpreted as exotic neutrino physics. 

In Ref. \cite{Giunti:2007xv}, the MiniBooNE anomaly was explained through a renormalization of the absolute neutrino flux and a simultaneous disappearance of electron neutrinos oscillating into sterile neutrinos (with $P_{\nu_e \to \nu_e} = 0.64 ^{+0.08}_{-0.07}$). The LSND and Gallium radioactive source experiment \cite{Hampel:1997fc, Abdurashitov:1998ne, Abdurashitov:2005ax} anomalies have been studied in Ref. \cite{Giunti:2006bj}, where is it claimed that all these anomalies could be interpreted as an indication of the presence of, at least, one sterile neutrino with rather large mass (few eV's). Ref. \cite{Acero:2007su} also studied the compatibility of the Gallium results with the Bugey \cite{Declais:1994su} and Chooz \cite{Apollonio:2002gd} reactor experimental data, concluding that such a sterile neutrino should have a mass between one and two eV's. Finally, the MiniBooNE collaboration performed global fits of MiniBooNE, LSND, KARMEN2, and Bugey experiments in presence of a fourth sterile neutrino \cite{AguilarArevalo:2008sk} (assuming no renormalization issue for MiniBooNe unlike Ref.~\cite{Giunti:2007xv}). When all four experiments are combined, the compatibility between them is found to be very low (4\%); however, when only three of them are included, the compatibility level is usually reasonable (the largest tension being found between LSND and Bugey). In this analysis, the preferred value of the sterile neutrino is usually smaller than 1eV, but still of possible cosmological relevance (for instance, for all four experiments, the best fit corresponds to $\Delta m^2 \sim 0.2 - 0.3$~eV$^2$).

These various developments suggest that it is important to scrutinize cosmological bounds on scenarios with one light sterile neutrino, which could help ruling them out, given that current bounds on the total neutrino mass assuming just three active neutrinos are as low as $\sum m_{\nu} < 0.61$eV (using WMAP5, BAO and SN data \cite{Komatsu:2008hk}). This result cannot be readily applied to the models which we consider here. Indeed, scenarios with extra neutrinos require a specific cosmological analysis, for the simple reason that besides affecting the total neutrino mass, additional neutrinos also increase the abundance of relativistic particles in the early universe.

From the point of view of Cosmology, there have been many works constraining simultaneously the sum of neutrino masses and the contribution to the relativistic energy density component of the Universe, parametrized as the effective number of neutrinos, $N_{\text{eff}}$ (see for example \cite{Dolgov:2002wy, Crotty:2004gm, Cuoco:2005qr, Lesgourgues:2006nd, Hannestad:2006mi, Bell:2005dr}). Most of these works assume either that the heaviest neutrino (and hence the most relevant one from the point of view of free-streaming) has a thermal distribution, sharing the same value of temperature as active neutrinos, or that all neutrinos are degenerate in mass. However, the results of Refs.~\cite{Cirelli:2006kt, Dodelson:2005tp} can also be
applied to the case of very light active neutrinos plus one heavier, non-necessarily thermal sterile neutrino, which is the most interesting case for explaining oscillation anomalies. In terms of physical motivations, it is very likely that the light sterile neutrino required by the LSND anomaly acquires a thermal distribution in the early universe, through oscillations with active neutrinos in presence of a large mixing angle \cite{DiBari:2001ua}. On the other hand, there are some proposals to avoid these contrains (for a list of some scenarios, see \cite{Abazajian:2002bj}). One of such possibilities is based on a low reheating temperature ($T_R$) Universe \cite{Giudice:2000dp, Gelmini:2004ah, Gelmini:2008fq, Yaguna:2007wi}, in which, for a sufficiently low $T_R$, the sterile neutrinos could be non-thermal \cite{Kawasaki:2000en} and its production would be suppressed \cite{Gelmini:2008fq}, such that usual cosmological bounds are evaded. In fact, in these models, sterile neutrinos are allowed to have a large mass without entering in conflict with other experimental results, while $T_{\text{dec}} \lesssim T_R \lesssim 10$MeV ($T_{\text{dec}}$ being the temperature of the cosmic plasma at neutrino decoupling).

In absence of thermalization, cosmological bounds on the sterile neutrino mass become potentially weaker. Hence, it is interesting to study the compatibility of recently proposed scenarios with a light sterile neutrino with the most recent cosmological data, keeping in mind the possibility of a non-thermal distribution. The goal of this paper is hence to study the compatibility of cosmological experimental data (WMAP5 plus small-scale CMB data, SDSS LRG data, SNIa data from SNLS and conservative Lya data from VHS) with the hypothesis of a sterile neutrino with the characteristics sketched above, i.e., with a mass roughly of the order of the electron-Volt, and a contribution to $N_{\text{eff}}$ smaller than one.

\section{Light sterile neutrino in cosmology: physical effects and parametrization}
\label{theory}

If a population of free-streaming particles becomes non-relativistic after photon decoupling, its physical effects on the cosmological background and perturbation evolution are mainly described by three quantities: \begin{enumerate}
\item
its contribution to the relativistic density before photon decoupling, which affects the redshift of radiation/matter equality, usually parametrized by an effective neutrino number (standing for the relativistic density of the species divided by that of one massless neutrino family in the instantaneous decoupling (id) limit):
\begin{equation}
\Delta N_{\rm eff} \equiv \frac{\rho_s^{\rm rel}}{\rho_\nu}
= \left[
{\frac{1}{\pi^2} \int \!\! dp \,\, p^3 f(p)} 
\right] /
\left[{\frac{7}{8} \frac{\pi^2}{15}
{T_\nu^{\rm id}}^4} \right]
\end{equation}
with $T_\nu^{\rm id} \equiv (4/11)^{1/3} T_{\gamma}$,
\item
its current energy density, which affects (i) the current energy budget of the Universe (with various consequences for the CMB and LSS spectra, depending on which other parameters are kept fixed), and (ii) the amplitude reduction in the small-scale matter power spectrum due to these extra massive free-streaming particles, parametrized by the dimensionless number $\omega_s$:
\begin{equation}
\omega_s \equiv \Omega_s h^2 = \left[
{\frac{m}{\pi^2} \int \!\! dp \,\, p^2 f(p)} 
\right] \times \left[\frac{h^2}{\rho_c^0}\right]
\end{equation}
where $\rho_c^0$ is the critical density today and $h$ the reduced Hubble parameter,
\item
the comoving free-streaming length of these particles when they become non-relativistic, which controls the scale at which the suppression of small-scale matter fluctuations occurs. This length can easily be related to the average velocity of the particles today, $\langle v_s \rangle$\footnote{The minimum comoving free-streaming wavenumber $k_{\rm fs}$ is controlled by $\Omega_m$ and by the ratio $a(t_{\rm nr})/\langle v_s(t_{\rm nr}) \rangle$ evaluated when $T=m$, i.e. when $a(t_{\rm nr}) \sim \langle v_s(t_0) \rangle a(t_0)$.
Given that $\langle v_s(t_{\rm nr}) \rangle \sim \langle v_s(t_0) \rangle a(t_0)/a(t_{\rm nr})$, the minimum comoving free-streaming length just depends on $\langle v_s(t_{0}) \rangle$ and $\Omega_m$.}.
\end{enumerate}
However, for whatever assumption concerning the phase-space distribution function $f(p)$, the three numbers ($\Delta N_{\rm eff}$, $\omega_s$, $\langle v_s \rangle$) satisfy  a constraint equation. Indeed, the average velocity of the particles today (assumed to be in the non-relativistic regime) is given {\it exactly} by
\begin{equation}
\langle v_s \rangle \equiv 
\frac{\int p^2 d p \, \frac{p}{m} f(p)}
{\int p^2 d p \, f(p)} 
 = \frac{7}{8} \frac{\pi^2}{15} \left(\frac{4}{11} \right)^{4/3}
\frac{T_{\rm CMB}^4 h^2}{\rho_c} \frac{\Delta N_{\rm eff}}{\omega_s}
= 5.618 \times 10^{-6} \frac{\Delta N_{\rm eff}}{\omega_s}
\end{equation}
in units where $c=k_B=\hbar=1$, and taking $T_{\rm CMB}=2.726$K. Hence, the three physical effects described above depend on only {\it two independent parameters}.

Reducing the physical impact of any population of massive free-streaming particles to these three effects (and two independent parameters) is a simplification: two models based on different non-thermal phase-space distributions $f(p)$ can in principle share the same numbers ($\Delta N_{\rm eff}$, $\omega_s$, $\langle v_s \rangle$) and impact the matter power spectrum differently. Indeed, the free-streaming effect depends on the details of $f(p)$ (including high statistical momenta like $\int \!\! dp \,\, p^4 f(p)$, etc.) However, the conclusions of Ref. \cite{Cuoco:2005qr} indicate that for many models with non-thermal distorsions, observable effects can indeed be parametrized by two combinations of ($\Delta N_{\rm eff}$, $\omega_s$, $\langle v_s \rangle$) with good accuracy: other independent parameters would be very difficult to observe\footnote{This conclusion does not apply when the non-thermal distribution $f(p)$ has a sharp peak close to $p=0$. In this case, particles with very small momentum should be counted within the CDM component, not within the extra massive free-streaming component. Otherwise, one would obtain values of $\omega_s$ and $\langle v_s \rangle$ based on an averages between cold and hot/warm particles; then, these parameters would not capture the correct physical effects (see \cite{Boyarsky:2008prep})}.

Let us compute the three parameters ($\Delta N_{\rm eff}$, $\omega_s$, $\langle v_s \rangle$) for simple cases. For one species of thermalized free-streaming particles with mass $m_s$, sharing the same temperature as active neutrinos in the instantaneous decoupling limit, one gets:
\begin{equation}
\Delta N_{\rm eff}=1, \qquad
\omega_s = \frac{m_s}{94.05~{\rm eV}}, \qquad
\langle v_s \rangle =
\frac{7 \pi^4}{180 \zeta(3)} \frac{T_\nu^{\rm id}}{m_s}=
\frac{0.5283~{\rm meV}}{m_s}.
\end{equation}
For a light thermal relic with a Fermi-Dirac distribution and a
different temperature $T_s$,
these quantities become
\begin{equation}
\Delta N_{\rm eff} = \left( \frac{T_s}{T_{\nu}^{\rm id}} \right)^4, \qquad
\omega_s = \frac{m_s}{94.05 {\rm eV}} \left(\frac{T_s}{T_{\nu}^{\rm id}}\right)^3, \qquad 
\langle v_s \rangle = \frac{0.5283~{\rm meV}}{m_s} \left(\frac{T_s}{T_{\nu}^{\rm id}}\right)~.
\end{equation}
For a non-thermal relic with a free function $f(p)$, there is an
infinity of possible models. A popular one is the Dodelson-Widrow
scenario \cite{Dodelson:1993je} (also referred as the ``non-resonant
production scenario''), motivated by early active-sterile neutrino
oscillations in the limit of small mixing angle and zero leptonic
asymmetry, which corresponds to the phase-space distribution
\begin{equation}
f(p)=\frac{\chi}{e^{p/T_{\nu}}+1} \label{lntsn_f2}
\end{equation}
where $\chi$ is an arbitrary normalization factor. In this case,
in the approximation $T_{\nu}=T_{\nu}^{\rm id}$,
the three ``observable'' parameters read
\begin{equation}
\Delta N_{\rm eff} = \chi, \qquad
\omega_s = \frac{m_s}{94.05 {\rm eV}} \chi, \qquad
\langle v_s \rangle = \frac{0.5283~{\rm meV}}{m_s}~.
\end{equation}
Hence, a Dodelson-Widrow (DW) model shares that same ``observable''
parameters ($\Delta N_{\rm eff}$, $\omega_s$, $\langle v_s \rangle$)
as a thermal model with $m_s^{\rm thermal}= m_s^{\rm DW} \chi^{1/4}$
and $T_s=\chi^{1/4} T_{\nu}$. Actually, for these two models, the
degeneracy is exact: it can be shown by a change of variable in the
background and linear perturbation equations that the two models are strictly
equivalent from the point of view of cosmological observables
\cite{Colombi:1995ze,Cuoco:2005qr}.  As mentioned before, in the general
case, two models sharing the same ($\Delta N_{\rm eff}$, $\omega_s$,
$\langle v_s \rangle$) are not always strictly equivalent, but can be thought
to be hardly distinguishable even with future cosmological data.  For
instance, the low-temperature reheating model analyzed in
\cite{Gelmini:2004ah, Gelmini:2008fq} leads to a distribution of the form
\begin{equation}
f(p)=\frac{\chi p }{e^{p/T_{\nu}}+1}~. \label{lntsn_f3}
\end{equation}
This model would in principle deserve a specific analysis, but in
good approximation we can expect that by only exploring the parameter
space of thermal models (or equivalently, of DW models), we will
obtain some very generic results, covering in good
approximation most possibilities for the non-thermal distorsions.

\section{Analysis}

\subsection{Data}

In the following sections, we will present the results of various runs
based on the Boltzmann code CAMB \cite{Lewis:1999bs} and cosmological
parameter extraction code CosmoMC \cite{Lewis:2002ah}. We modified
CAMB in order to implement the proper phase-space distribution $f(p)$
of the thermal or DW model. For simplicity, we assumed in all runs
that the three active neutrinos can be described as massless
particles. In order to obtain a Bayesian probability distribution for
each cosmological parameters, we ran CosmoMC with flat priors on the
usual set of six parameters $\omega_b$, $\omega_{\rm dm}=
\omega_s+\omega_{\rm cdm}$, $\theta$, $\tau$, $A_s$, $n_s$ (see
e.g. \cite{Spergel:2006hy}), plus two extra parameters describing the
sterile neutrino sector, that will be described in the next sections.
We choose the following data set: WMAP5 \cite{Dunkley:2008ie} plus
small-scale CMB data (ACBAR \cite{Reichardt:2008ay}, CBI
\cite{Sievers:2005gj}, Boomerang \cite{MacTavish:2005yk}), the galaxy
power spectrum of the SDSS LRG \cite{Tegmark:2006az} with flat prior
on $Q$ \cite{Hamann:2007pi,Hamann:2008we}, SNIa data from SNLS
\cite{Astier:2005qq} and conservative Lyman-$\alpha$ data from VHS
\cite{Viel:2004bf}. We do not include more recent Lyman-$\alpha$ data
sets, which have much smaller errorbars, but for which the
deconvolution of non-linear effects depends on each particular
cosmological model, and requires specific hydrodynamical simulations.

\subsection{General analysis \label{GEN}}

Our first goal is to obtain simple results with a wide range of
applications. Hence, we should not parametrize the effect of sterile
neutrinos with e.g. their mass or temperature: in that case, our
results would strongly depend on underlying assumptions for $f(p)$.
It is clear from section \ref{theory} that nearly ``universal''
results can be obtained by employing two combinations of the
``observable parameters'' $\Delta N_{\rm eff}$, $\omega_s$ and
$\langle v_s \rangle$ (and eventually of other parameters of the
$\Lambda$CDM model). Here we choose to vary the current dark matter
density fraction $f_s=\omega_s/(\omega_{s}+\omega_{\rm cdm})$ and the
current velocity dispersion $\langle v_s \rangle$. As will be clear
from our results, these two parameters capture the dominant observable
effects, and lead to very clear bounds, since their correlation with
other $\Lambda$CDM model parameters is insignificant.  Our limits on
$f_s$ and $\langle v_s \rangle$ apply exactly to the thermal case and
DW case, and approximately to most other cases (modulo the caveat
described in the second footnote of section \ref{theory}).

\begin{figure}[t!]
\includegraphics[angle=-90,width=14cm]{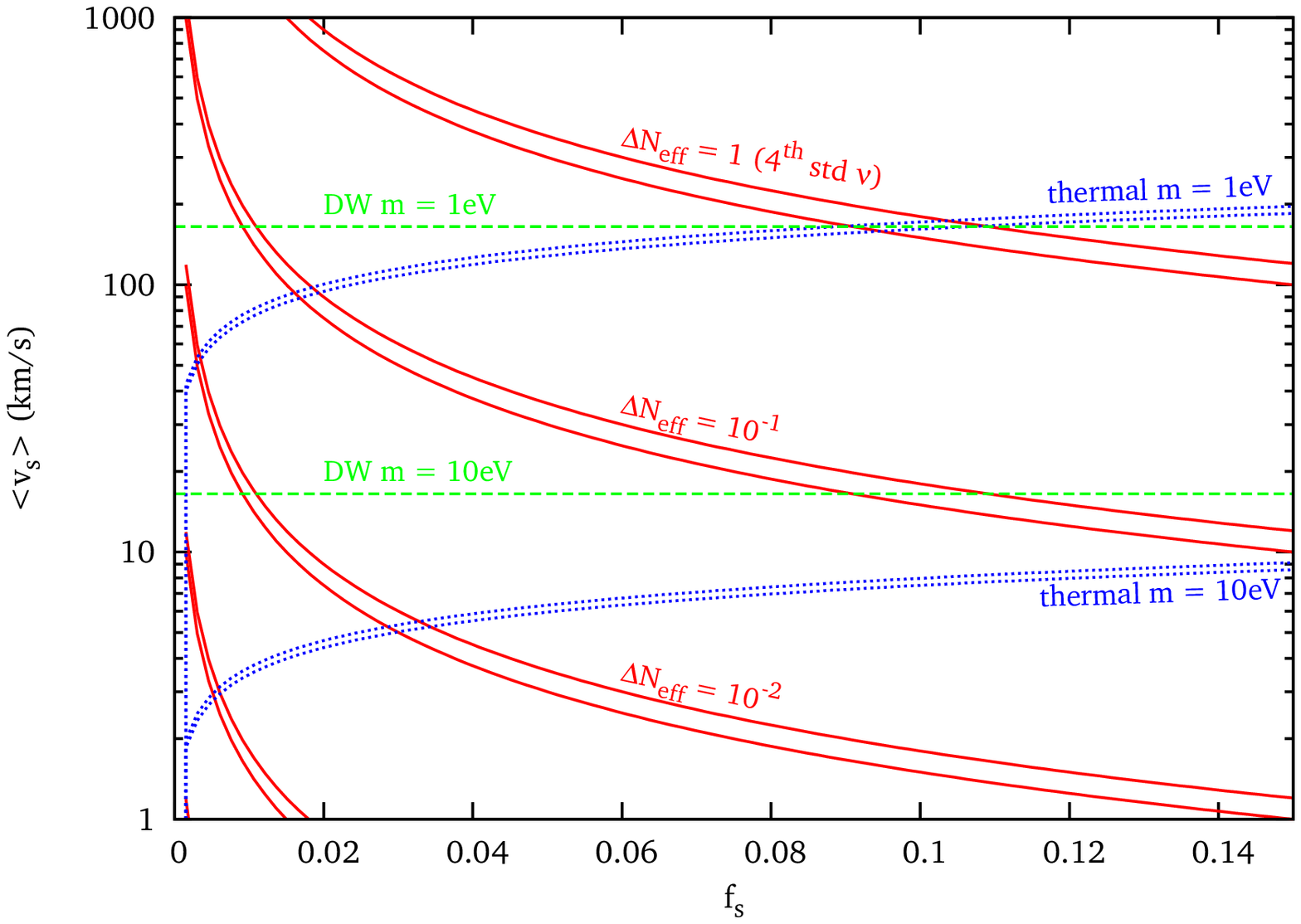}
\includegraphics[angle=-90,width=14cm]{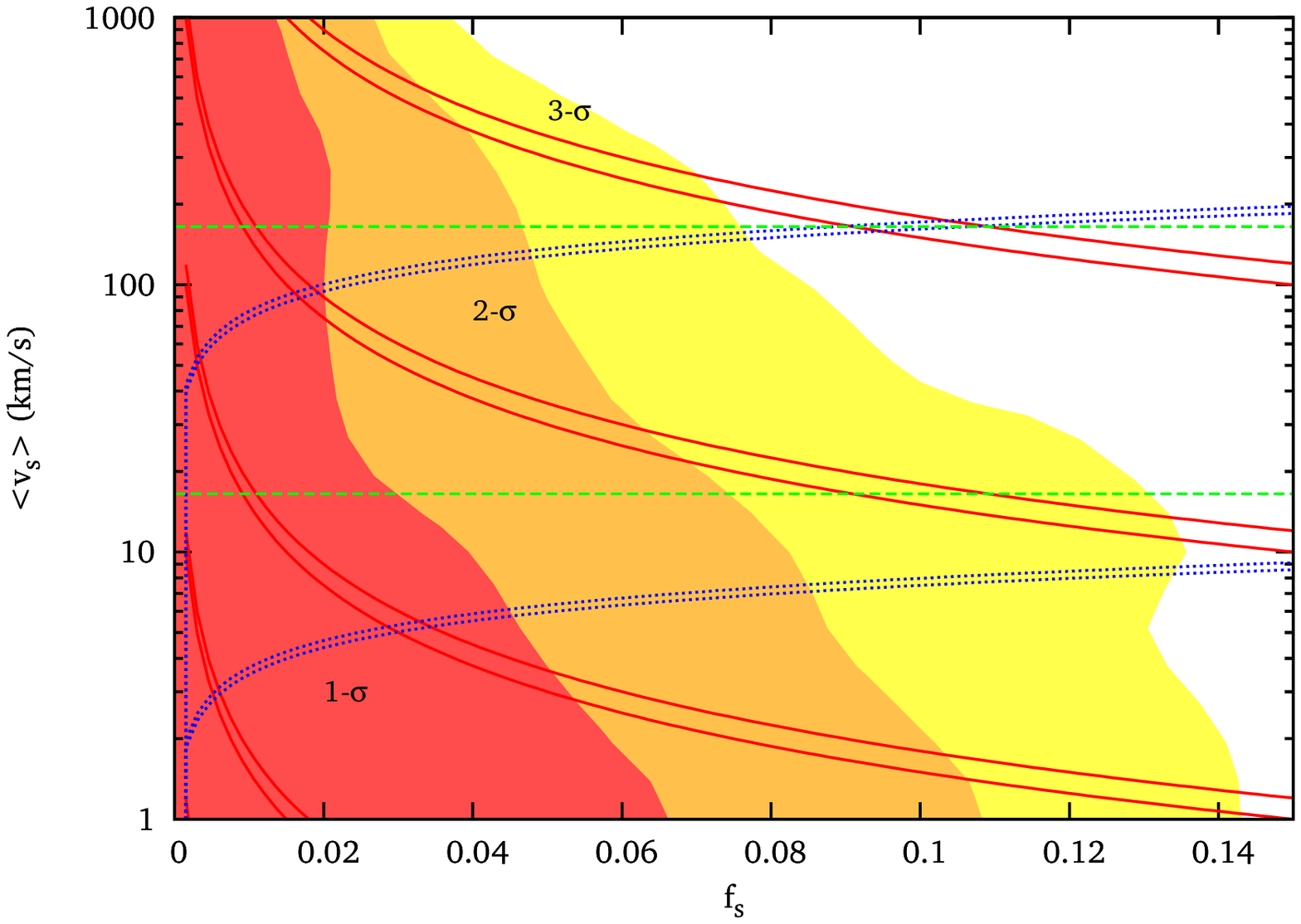}
  \caption{{\it (Top)} the parameter space ($f_s$,$\langle v_s
    \rangle$) chosen in our general analysis. The thin bands delimited
    by red/solid lines show regions of equal $\Delta N_{\rm eff}$
    (assuming $\omega_{\rm dm} = 0.11 \pm 0.01$); these bands are
    fully model-independent. We also show the model-dependent regions
    of equal mass, delimited by blue/dotted lines for the case of
    early decoupled thermal relics, and consisting in horizontal
    green/dashed lines for Dodelson-Widrow sterile neutrinos. {\it
    (Bottom)} same with, in addition, the regions allowed at the
    68.3\% (1$\sigma$), 95.4\% (2$\sigma$) and 99.7\% (3$\sigma$) C.L.
    by our cosmological data set, in a Bayesian analysis with flat
    priors on $f_s$ and $\log_{10}\langle v_s \rangle$ within the
    displayed range.  }\label{figA}
\end{figure}

Our parameter space is represented in Figure~\ref{figA}. We adopt
a logarithmic scale for $\langle v_s \rangle$ and display the 
interesting range
\begin{equation}
1~{\rm km/s} < \langle v_s \rangle < 1000~{\rm km/s}~.
\end{equation}
Indeed, with out dataset, particles with smaller velocities would be
indistinguishable from cold dark matter; instead, particles with
larger velocities would either have $\Delta N_{\rm eff} > 1$ (a case
beyond the motivations of this work, and anyway very constrained by
the data) or $f_s < 0.02$ (being indistinguishable from extra
relativistic degrees of freedom).  Assuming a particular value for
$\omega_{\rm dm}=\omega_s+\omega_{\rm cdm}$ and for $\Delta N_{\rm
eff}$, it is possible to compute the velocity dispersion $\langle v_s
\rangle$ as a function of $f_s$.  Since the CMB and LSS data give
precise constraints on $\omega_{\rm dm}$, regions of equal $\Delta
N_{\rm eff}$ correspond to thin bands in the ($f_s$, $\langle v_s
\rangle$) plane. We show these bands in Figure~\ref{figA} for $10^{-3}
< \Delta N_{\rm eff} < 1$ under the assumption that $\omega_{\rm dm} =
0.11 \pm 0.01$, which corresponds roughly to the 95\% confidence limits
(C.L.) from all our runs. These iso-$\Delta N_{\rm eff}$ bands are
completely model-independent.

Instead, regions of equal mass can only be plotted for a particular
model. In Figure~\ref{figA}, we show the bands corresponding to
$m=$1~eV and 10~eV, either in the case of early decoupled
thermal relics (blue/dotted lines) or in the DW case (green/dashed lines).  For any
given mass, these bands intersect each other in a location
corresponding to the case of one fourth standard neutrino species with
$\Delta N_{\rm eff}=1$.

We ran CosmoMC with top-hat priors on $f_s$ (in the physical range
$[\,0,1]$) and on $\log_{10}[{\langle v_s \rangle} / {1 {\rm km/s}}]$
(in the range $[0,3]$ motivated by the previous discussion).
Our results are summarized in Figure~\ref{figA} (bottom).
We see that the upper bound on $f_s$ decreases smoothly as the
velocity dispersion increases: when the particles have a larger
velocity dispersion, their free-streaming wavelength is larger, so the
step-like suppression in the power spectrum (which amplitude depends on
$f_s$) is more constrained. For $\langle v_s \rangle \sim 1~{\rm
  km/s}$, we find $f_s \lesssim 0.1$ at the $2\sigma$ C.L., while for $\langle
v_s \rangle \sim 100~{\rm km/s}$, we find $f_s \lesssim 0.06$ at the $2\sigma$
C.L. When the velocity dispersion becomes larger than $100~{\rm
  km/s}$, the upper bound on $f_s$ decreases even faster as a function
of $\langle v_s \rangle$. This is the case of a HDM component with
significant contribution to the number of relativistic d.o.f., for
which the observational bounds derive from a combination of the first
and second effects described in section \ref{theory}: in this limit,
in addition to being sensitive to the free-streaming effect, the data
disfavor a significant increase of the total radiation density
corresponding to $\Delta N_{\rm eff}$ of order one or larger.

\begin{figure}[t!]
\begin{center}
\includegraphics[angle=-90,width=14cm]{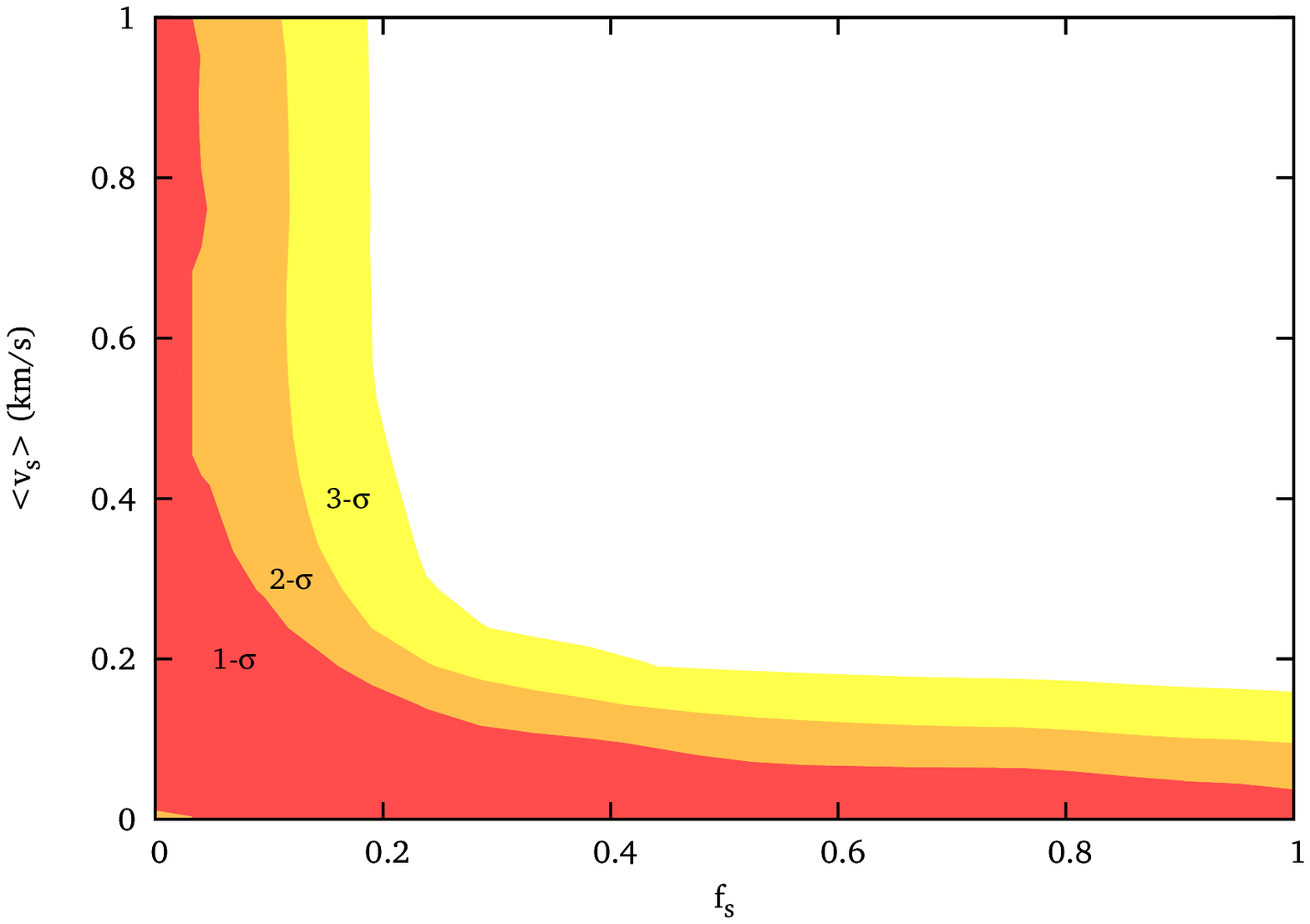}
  \caption{1$\sigma$, 2$\sigma$ and 3$\sigma$ contours of the marginalized
    likelihood for the two parameters ($f_s$, $\langle v_s \rangle$),
    with different priors than in previous figures. As explained in the text, this plot shows
    the region where the sterile neutrino is heavy and behaves like
    warm dark matter, in complement to Figure 1, which is based on a
    different range/prior for $\langle v_s \rangle$ adapted to the
    case of a light, hot sterile neutrino.}\label{fig_f_v_col}
\end{center}
\end{figure}

We should stress that the details of our results depend on the
underlying priors.  For instance, one could use a flat prior on
$\langle v_s \rangle$ instead of its logarithm.  Running in the range
$0 < \langle v_s \rangle < 1000~{\rm km/s}$ with such a prior would
give more focus on the large-$\langle v_s \rangle$ allowed region of
Figure~\ref{figA}. However, it would be more interesting to focus on small
velocities, in order to understand how our results can be extended without
any discontinuity to the
case of warmer and heavier dark matter.  For this
purpose, we ran CosmoMC with a top-hat prior on $0 < \langle v_s \rangle <
1~{\rm km/s}$, and obtained the results shown in
Figure~\ref{fig_f_v_col}. These results are identical to those
published in Reference~\cite{Boyarsky:2008xj} (figure 7). By gluing
figure~\ref{figA} on top of \ref{fig_f_v_col}, one can obtain a full
coverage of the parameter space of $\Lambda$CDM models completed by
one extra (hot or warm) dark matter species.  Figure \ref{fig_f_v_col}
shows the transition from the region in which this extra species is
indistinguishable from cold dark matter (when $\langle v_s \rangle
\leq 0.1~{\rm km/s}$, the fraction $f_s$ is unconstrained) to the
region in which it is warm (for $0.4 \leq \langle v_s \rangle \leq 1~{\rm
  km/s}$, there is a nearly constant bound $f_s \lesssim 0.1$ at the
2-$\sigma$ level).  Figure ~\ref{figA} shows instead the transition
from warm particles to hot particles (with velocities comparable to those
of active neutrinos). The two plots perfectly match each other along the  
$\langle v_s \rangle = 1~{\rm km/s}$ axis, on which the sterile neutrino fraction
is bounded by $f_s \lesssim 0.1$ (2-$\sigma$).


\subsection{Mass/temperature bounds in the thermal case \label{TH}}

\begin{figure}[t!]
\includegraphics[angle=-90,width=14cm]{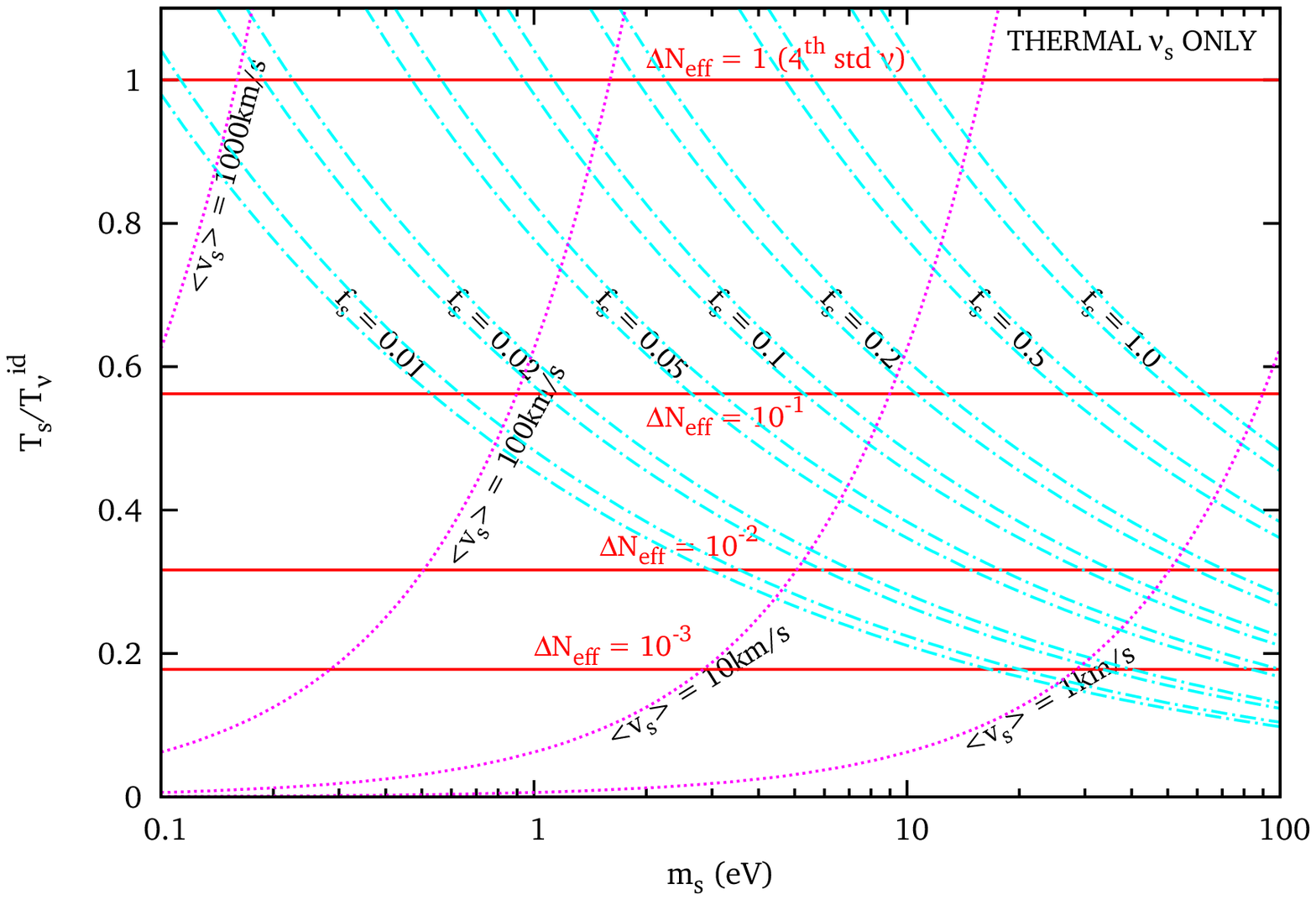}
\includegraphics[angle=-90,width=14cm]{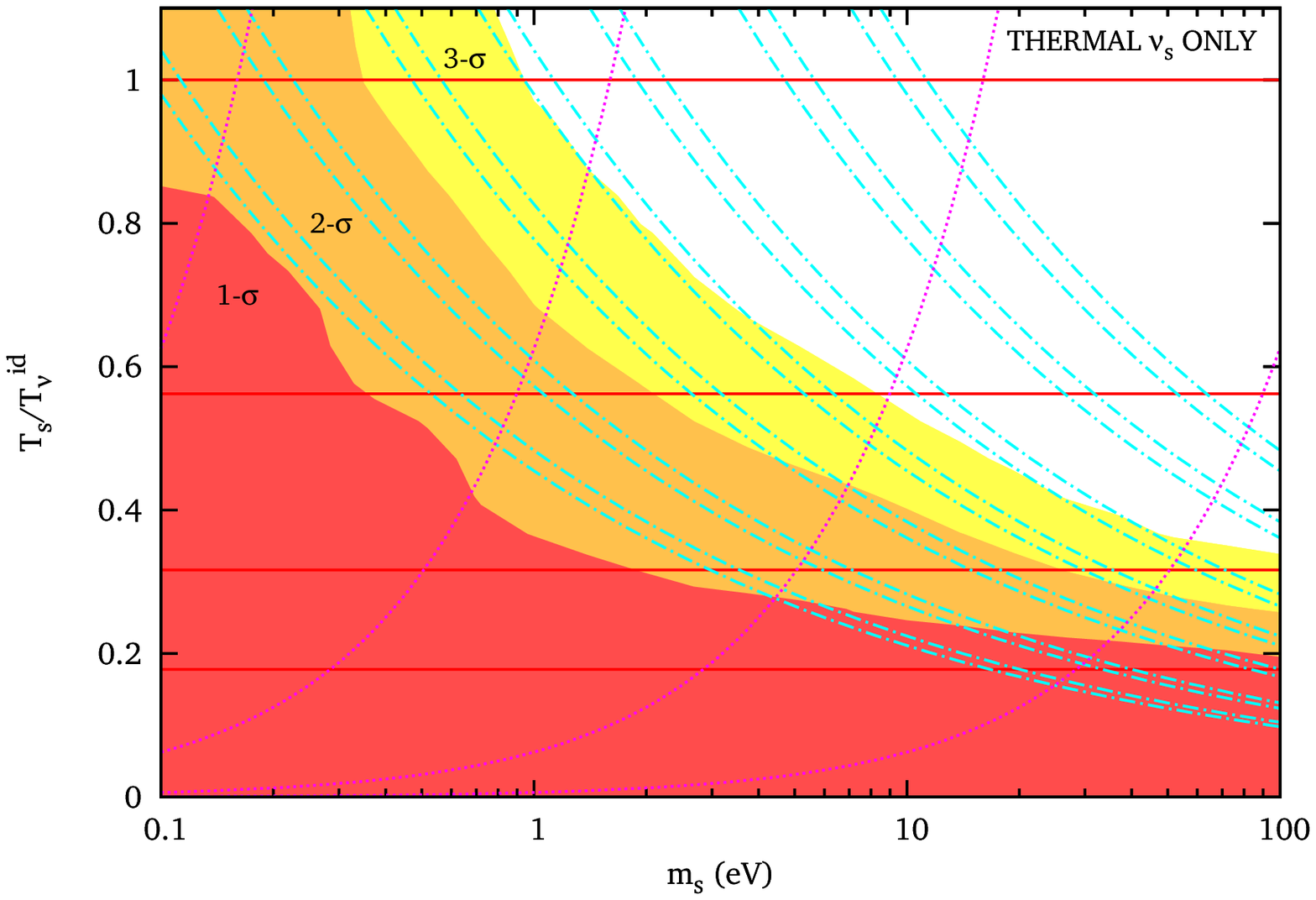}
  \caption{{\it (Top)} the parameter space ($m_s$,$T_s/T_\nu^{\rm
    id}$) used in the particular case of early decoupled thermal
    relics of temperature $T_s$ (with $T_\nu^{\rm id} \equiv
    (4/11)^{1/3} T_{\gamma}$). The thin bands delimited by
    blue/dot-dashed lines show regions of equal $f_s$ (assuming
    $\omega_{\rm dm} = 0.11 \pm 0.01$); the magenta/dotted lines
    correspond to fixed values of the velocity dispersion today;
    horizontal red/solid lines to fixed $\Delta N_{\rm eff}$. {\it
    (Bottom)} same with, in addition, the regions allowed at the
    68.3\% (1$\sigma$), 95.4\% (2$\sigma$) and 99.7\% (3$\sigma$) C.L.
    by our cosmological data set, in a Bayesian analysis with flat
    priors on $\log_{10}(m_s)$ and $T_s/T_\nu^{\rm id}$ within the
    displayed range. }\label{figB}
\end{figure}

We now focus on the particular case of early decoupled thermal relics, with a
Fermi-Dirac distribution and a temperature $T_s$. These models
can be parametrized by the
mass $m_s$ and the temperature in units of the neutrino
temperature, $T_s/T_\nu^{\rm id}$. Our parameter space -- and the
correspondence with the previous parameters $\Delta N_{\rm eff}$,
$f_s$, $\langle v_s \rangle$ -- is shown in Figure~\ref{figB}. In this
analysis, we want to focus again on light sterile neutrinos rather than WDM;
hence we are not interested in velocities smaller than 1~km/s
today. We are not interested either in the
case of enhanced particles with $\Delta N_{\rm eff} > 1$.  Then, as
can be checked in Figure~\ref{figB}, the ensemble of interesting
models can be covered by taking a top-hat prior on
$\log_{10}(m_s/1~{\rm eV})$ in the range $[-1,2]$, and on
$T_s/T_\nu^{\rm id}$ in the range $[0, 1]$.

The likelihood contours obtained for this case are shown in
Figure~\ref{figB} (bottom). They are consistent with our previous
results: when $\Delta N_{\rm eff} \sim 10^{-2}$
(and hence $T_s/T_\nu^{\rm id} \sim 0.3$), the upper bound on the
sterile neutrino fraction is $f_s<0.1$ at the $2\sigma$ C.L.; 
then this bound decreases smoothly when $T_s$ increases.
For a fourth standard neutrino
with $T_s = T_\nu^{\rm id}$, the $2\sigma$ C.L. (resp. $3\sigma$ C.L.) bound
is $m_s \lesssim 0.4$ eV (resp. 0.9~eV).

This figure can be conveniently used for model building: for a given
value of the mass, it shows what should be the maximal temperature of
the thermal relics in order to cope with cosmological observations;
knowing this information and assuming a particular extension of the
particle physics standard model, one can derive limits on the
decoupling time of the particle. For instance, for a mass of
$m_s=0.5$~eV one gets $T_s/T_\nu^{\rm id} \lesssim 0.9$; for
$m_s=1$~eV, $T_s/T_\nu^{\rm id} \lesssim 0.7$; while for $m_s=5$~eV,
$T_s/T_\nu^{\rm id} \lesssim 0.5$.  This figure can also be applied to
thermally produced axions, like in Refs.~\cite{Hannestad:2005df,Hannestad:2008js}.

\subsection{Mass bounds in the DW case \label{DW}}

\begin{figure}[t!]
\includegraphics[angle=-90,width=14cm]{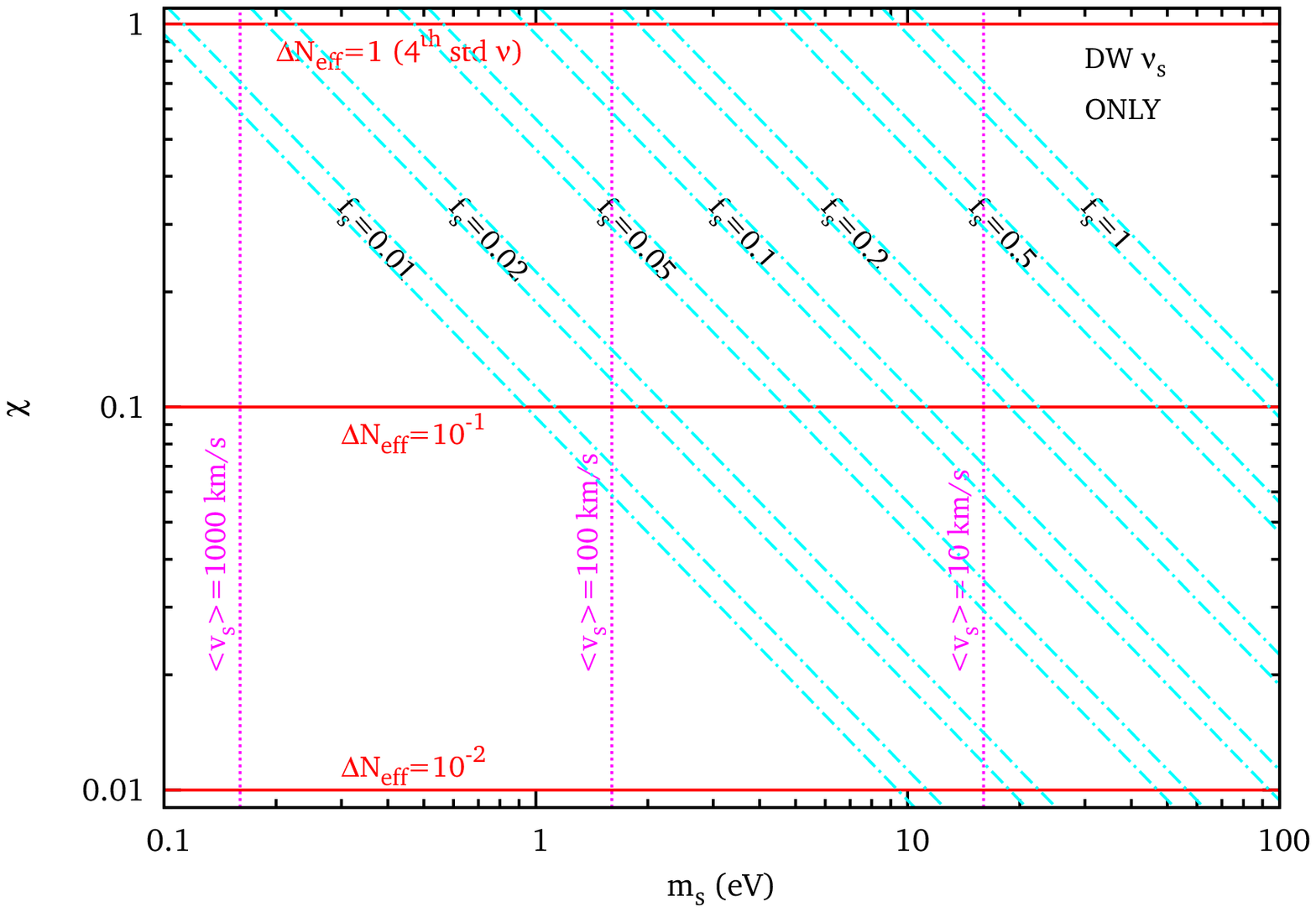}
\includegraphics[angle=-90,width=14cm]{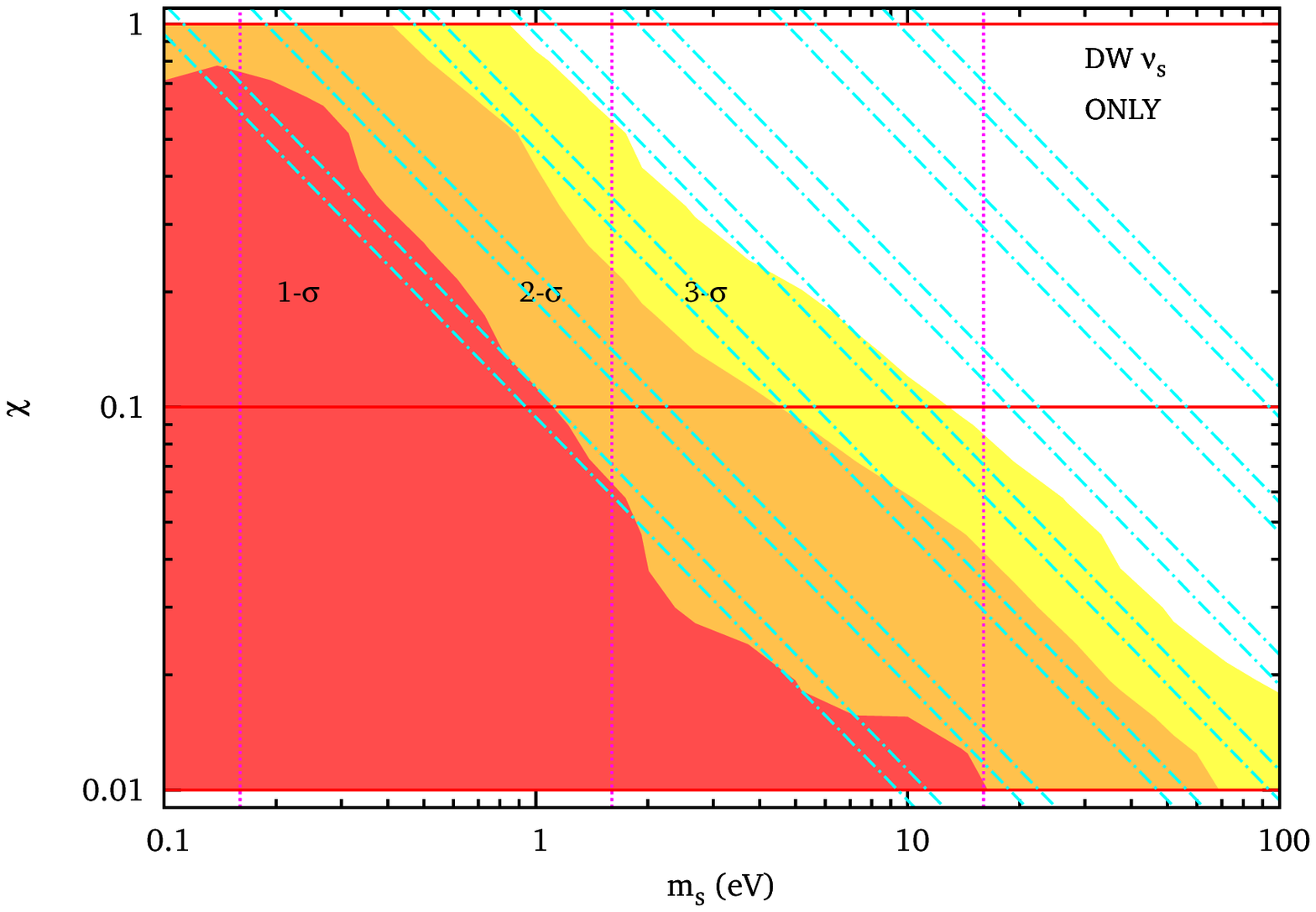}
  \caption{{\it (Top)} the parameter space ($m_s$,$\chi$) used in the
    particular case of DW relics. The thin bands delimited by
    blue/dot-dashed lines show regions of equal $f_s$ (assuming
    $\omega_{\rm dm} = 0.11 \pm 0.01$); the magenta/dotted lines
    correspond to fixed values of the velocity dispersion today;
    horizontal red/solid lines to fixed $\Delta N_{\rm eff}$.  {\it
      (Bottom)} same with, in addition, the regions allowed at the
    68.3\% (1$\sigma$), 95.4\% (2$\sigma$) and 99.7\% (3$\sigma$)
    C.L. by our cosmological data set, in a Bayesian analysis with
    flat priors on $\log_{10}(m_s)$ and $\log_{10}(\chi)$ within the
    displayed range.  }\label{figC}
\end{figure}

Finally, for Dodelson-Widrow relics with a distribution function equal
to that of standard neutrinos suppressed by a factor $\chi$ (which
is equal by definition to $\Delta N_{\rm eff}$), we can parametrize
the ensemble of models by $m_s$ and $\chi$.  Our
parameter space -- and the correspondence with $f_s$, $\langle v_s
\rangle$ -- is shown in Figure~\ref{figC}. Like in the previous section,
we are not interested in a current velocity dispersion smaller than
1~km/s today.  Then, as can be checked in Figure~\ref{figC}, the
ensemble of interesting models can be covered by taking a top-hat
prior on $\log_{10}(m_s/1~{\rm eV})$ in the range $[-1,2]$; in this
range, values of $\chi$ smaller than $10^{-2}$ would correspond to
tiny values of $f_s$, i.e. to particles indistinguishable from
massless particles; so, we can take a flat prior on
$\log_{10}(\chi)$ in the range $[-2, 0]$.

The likelihood contours obtained for this case are shown in
Figure~\ref{figC} (bottom). We are not surprised to find once more an
allowed region corresponding to $f_s \lesssim 0.1$ at the $2\sigma$ C.L. when
$\Delta N_{\rm eff}=\chi \sim 10^{-2}$ is negligible with respect to one, or less
when $\Delta N_{\rm eff}$ grows closer to one. For a fourth standard
neutrino with $T_s = T_\nu^{\rm id}$, the two definitions of the mass
(following from the thermal or from the DW cases) are equivalent, 
and indeed we find $m_s \lesssim 0.4$ eV ($2\sigma$ C.L.) or
$m_s \lesssim 0.9$ eV ($3\sigma$ C.L.)
like in section~\ref{TH}.

This figure can also be useful for model building: for a given value
of the mass, it shows what should be the maximal value of $\chi$
compatible with cosmological observations; in turn, this information
can be used to put bounds on the mixing angle between this relic and
active neutrinos in non-resonant production models {\it \`a la}
Dodelson \& Widrow.  For instance, for a mass of $m_s=1$~eV, the
$2\sigma$ C.L. gives $\chi \lesssim 0.5$; for $m_s=2$~eV, we get $\chi \lesssim
0.2$; while for $m_s=5$~eV, we get $\chi \lesssim 0.1$.

\subsection{Comparison with previous work}

The ensemble of cosmological models that we are exploring here is not
different from that studied by Dodelson, Melchiorri \& Slosar
\cite{Dodelson:2005tp} (called later DMS) or by Cirelli \& Strumia
\cite{Cirelli:2006kt} (called later CS); the difference between these
works and the present analysis consists in a different choice of
parameters, priors, data set, and also methodology in the case
of CS. 

For instance, Figure~6a of CS presents constrains in the space ($\log_{10}
\Delta N_{\rm eff}$, $\log_{10} m_s$) assuming a DW scenario.  Hence, their
parameter space is identical to the one we used in section~\ref{DW},
excepted for the prior range (which is wider in their case).  As far
as the data set is concerned, CS use some CMB and galaxy spectrum
measurements which are slightly obsolete by now; on the other hand,
they employ some additional information derived from BAO experiments,
and use SDSS Lyman-$\alpha$ data points that we conservatively
excluded from this analysis, since they assume a $\Lambda$CDM
cosmology. Finally, CS performed a frequentist analysis, and their bounds
are obtained by minimizing the $\chi^2$ over extra parameters (while in
the present Bayesian analysis, we marginalize over them given the priors).

\begin{figure}[t!]
\includegraphics[angle=-90,width=14cm]{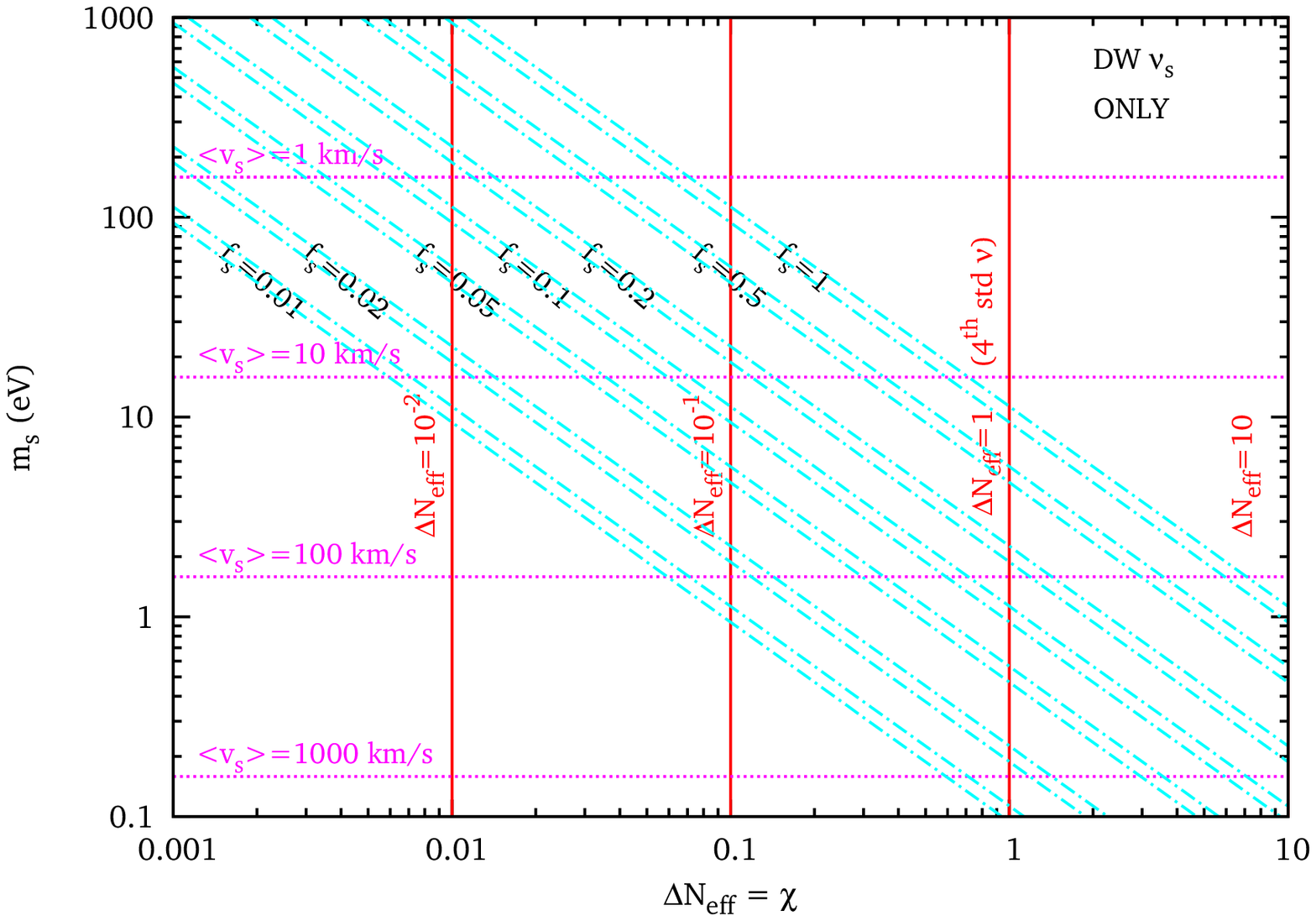}
\includegraphics[angle=-90,width=14cm]{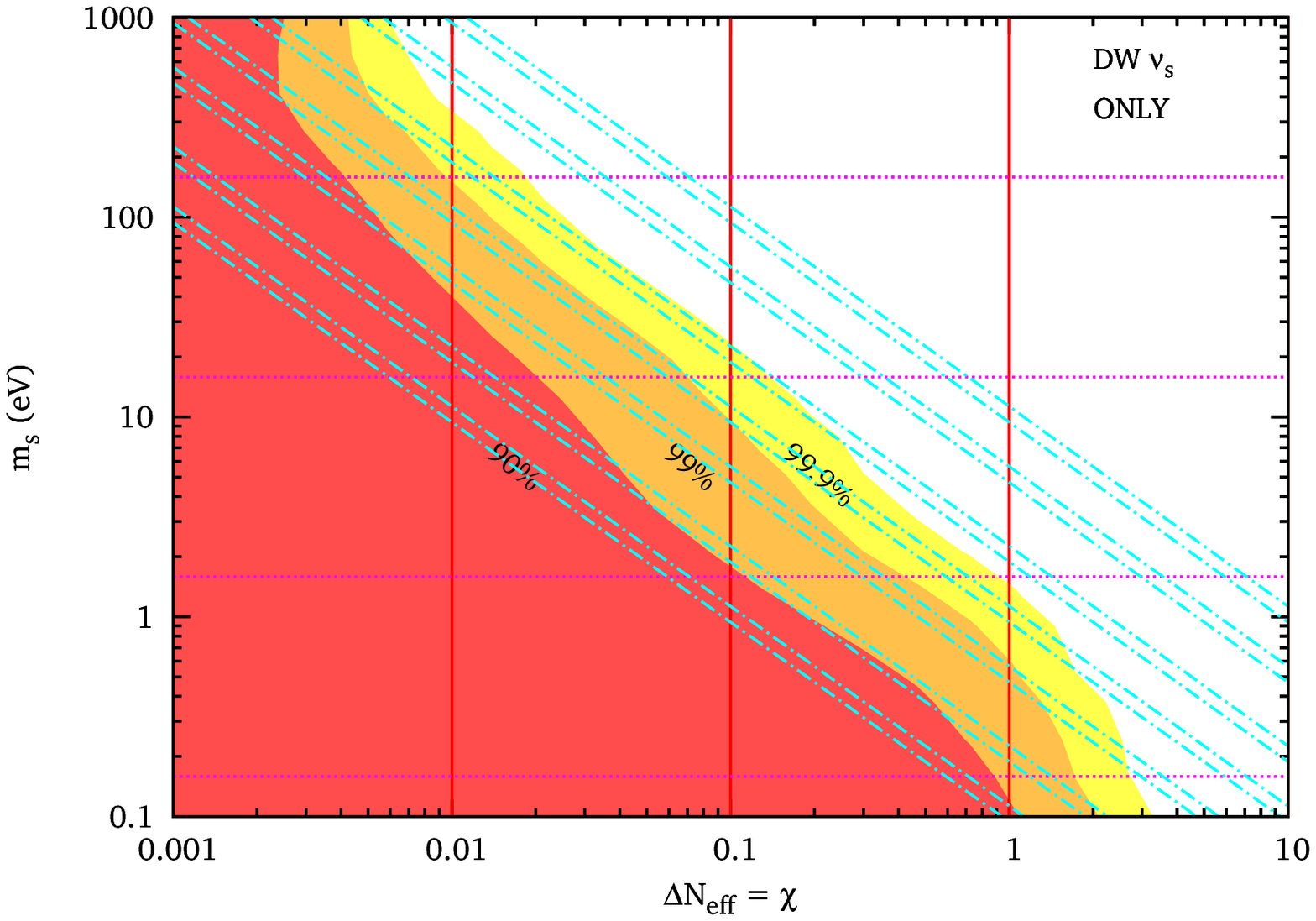}
  \caption{{\it (Top)} the parameter space ($\Delta N_{\rm
      eff}$,$m_s$) used for comparison with Cirelli \& Strumia in the
      particular case of DW relics. The thin bands delimited by
      blue/dot-dashed lines show regions of equal $f_s$ (assuming
      $\omega_{\rm dm} = 0.11 \pm 0.01$); the magenta/dotted lines
      correspond to fixed values of the velocity dispersion today;
      horizontal red/solid lines to fixed $\Delta N_{\rm eff}$.  {\it
      (Bottom)} same with, in addition, the regions allowed at the
      90\%, 99\% and 99.9\% C.L. by our cosmological data set, in a
      Bayesian analysis with flat priors on $\log_{10}(\Delta N_{\rm
      eff})$ and $\log_{10}(m_s)$ within the displayed range.
      }\label{figD}
\end{figure}

In order to compare our results with CS, we performed a run with
top-hat priors on $\log_{10} \chi = \log_{10} (\Delta N_{\rm eff})$ in
the range $[-3,1]$, and on $\log_{10} (m_s/1~{\rm eV})$ in the range
$[-1,3]$. In this particular run we compute the 90\%, 99\% and 99.9\%
C.L., following CS.
Our results are shown in Figure~\ref{figD}, and are consistent with
those of our general analysis. 

In spite of the different data set and methodology, the 90\% and 99\%
contours are found to be in very good agreement with CS in most of the
parameter space.  The major difference lies in the small mass region,
for which CS get more conservative limits on $\Delta N_{\rm eff}$ than
we do, and find a preference for non-zero values of the effective
neutrino number $0.5 < \Delta N_{\rm eff} < 4$ (at the 90\% C.L.).
This qualitative behavior has been nicely explained in
Refs.~\cite{Hamann:2007pi,Hamann:2008we}. It is due to the non-linear
corrections applied to the theoretical linear power spectrum before
comparing it with the observed SDSS and 2dF galaxy power spectra. The
approach used in this work (and in the default version of CosmoMC)
consists in marginalizing over a nuisance parameter $Q$ (describing
the scale-dependence of the bias) with a flat prior.  Instead,
following Ref.~\cite{Seljak:2006bg}, CS impose a gaussian prior on
$Q$. This results in biasing the results towards larger values of
$N_{\rm eff}$, and finding marginal evidence for $\Delta N_{\rm eff}
>0$. Of course, this assumption might turn out to be correct; however,
it is argued in Refs.~\cite{Hamann:2007pi,Hamann:2008we} that our
knowledge on $Q$ (based essentially on N-body simulations for some
particular cosmological models) is still too uncertain for getting
definite predictions.

\begin{figure}[t!]
\includegraphics[angle=-90,width=14cm]{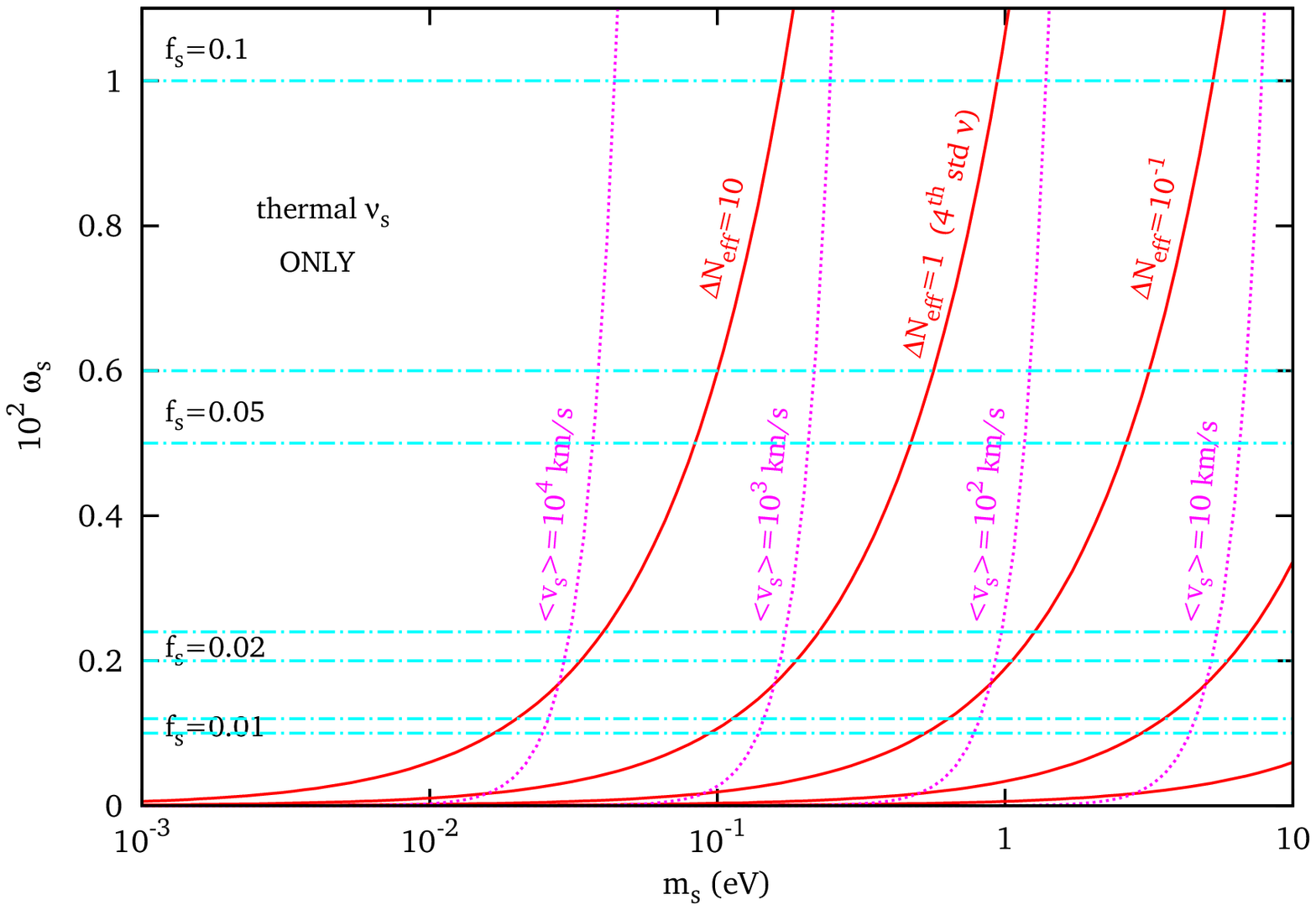}
\includegraphics[angle=-90,width=14cm]{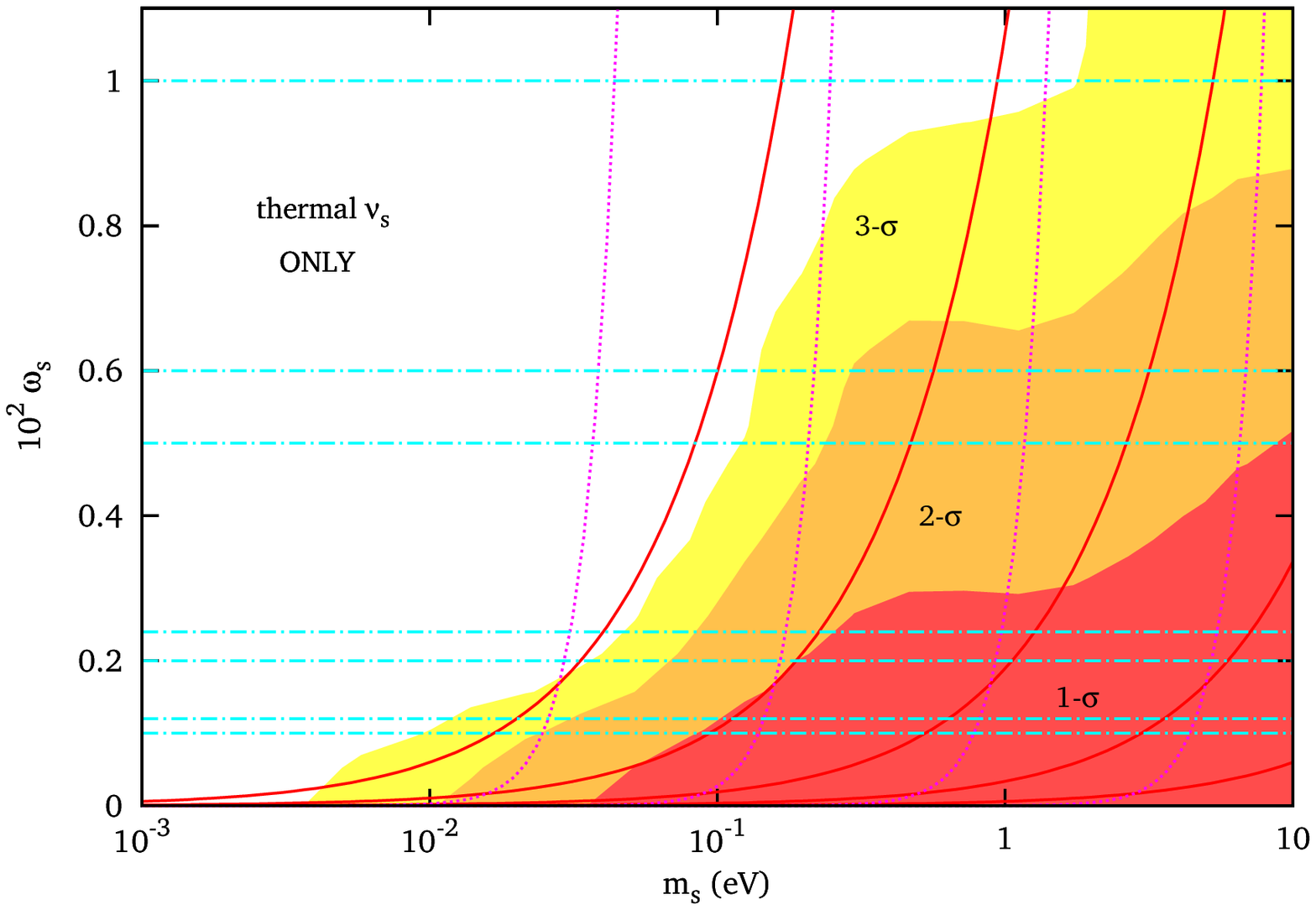}
  \caption{{\it (Top)} the parameter space ($m_s$,$\omega_s$)
    used for comparison with Dodelson, Melchiorri \& Slosar in the
    particular case of thermal relics. The thin bands delimited by
    blue/dot-dashed lines show regions of equal $f_s$ (assuming
    $\omega_{\rm dm} = 0.11 \pm 0.01$); the magenta/dotted lines
    correspond to fixed values of the velocity dispersion today;
    horizontal red/solid lines to fixed $\Delta N_{\rm eff}$.  {\it
      (Bottom)} same with, in addition, the regions allowed at the
    68.3\% (1$\sigma$), 95.4\% (2$\sigma$) and 99.7\% (3$\sigma$)
    C.L. by our cosmological data set, in a Bayesian analysis with
    flat priors on $\log_{10}(m_s)$ and
    $\omega_s$ within the displayed range.  }\label{figE}
\end{figure}

The analysis of DMS is Bayesian, like ours. The authors use top-hat priors
on the two parameters $-3<\log_{10}(m_s/1~{\rm eV})<1$ and $0 < \omega_s <
1$, roughly the same data set as CS, and employ the distribution function
of early decoupled thermal relics. Our results based on the same priors
(but a different data set) are shown in Figure~\ref{figE}, and are
consistent with the previous sections: at the 2$\sigma$ level, $\omega_s$
is such that $f_s \lesssim 0.1$ for $\Delta N_{\rm eff} \sim 10^{-2}$;
then, the bound on $f_s$ (and therefore on $\omega_s$) decreases smoothly
when $m_s$ decreases (and therefore $\langle v_s \rangle$ increases).
These results differ significantly from those of DMS, who find that the
upper bound on $\omega_s$ peaks near $m\sim0.25$~eV and then decreases
quickly. We do not observe such a behavior: our upper bound on $\omega_s$
increases (not so smoothly, but still monotonically) when $m_s$ increases,
in agreement with all previous results in this paper. This difference is
most likely due to the use made by DMS of more aggressive Lyman-$\alpha$
data from SDSS, of different galaxy power spectrum data, and of a prior on
$Q$, as in CS. This data set puts stronger limits on a possible suppression
of the small scale power spectrum. Actually, in absence of sterile
neutrinos, the same combination of data is known to produce very strong
bounds on neutrino masses, and to prefer $\Delta N_{\rm eff}$ slightly
larger than one \cite{Seljak:2006bg}; in presence of light sterile
neutrinos, the results of DMS show that this data also imposes a strong
bound $\omega_s < 0.001$ for $1 \, {\rm eV} < m_s < 10 \, {\rm eV}$, due to its
sensitivity to the sterile neutrino free-streaming effect. Our large scale
structure data set (conservative Lyman-$\alpha$ data from VHS, SDSS-LRG
and flat prior on $Q$) is not able to exclude this region.


\section{Conclusions}

In this work, we studied the compatibility of cosmological experimental data
with the hypothesis of a non-thermal sterile neutrino with a mass in
the range $0.1-10$ eV (or more), and a contribution to
$N_{\text{eff}}$ smaller than one. We computed Bayesian confidence
limits on different sets of parameters, adapted to the case of thermal
relics (section \ref{TH}), of non-resonantly produced sterile neutrinos 
{\it \`a
  la} Dodelson \& Widrow
(DW, section \ref{DW}),
or of generic parameters leading to nearly model-independent results (section \ref{GEN}).
In each case, we performed a specific parameter extraction from scratch, 
in order to obtain reliable results assuming flat priors on the
displayed parameters. For simplicity, we assumed that the masses of the three active neutrinos
are negligible with respect to that of the sterile neutrino.

For a cosmological data set consisting in recent CMB and LSS data, as well as older but
very conservative Lyman-$\alpha$ data, 
we found the conditional probability e.g. on the mass of a thermal relic given its temperature,
or on the mass of a DW neutrino given its density suppression factor, etc. These proabilities are such
that if the fourth neutrino is a standard one (with $\Delta N_{\rm eff}=1$), it should have a mass
$m_s \lesssim 0.4$ eV ($2\sigma$ C.L.) or
$m_s \lesssim 0.9$ eV ($3\sigma$ C.L.).

At the $3\sigma$ C.L., a mass $m_s = 1$ eV can be accommodated with the
data provided that this neutrino is thermally distributed with
$T_s/T_\nu^{\rm id} \lesssim 0.97$, or non-resonantly produced with $\Delta N_{\text{eff}}
\lesssim 0.9$. The bounds become dramatically tighter when the mass
increases.  At the same confidence level, a mass of just $m_s = 2$~eV
requires either $T_s/T_\nu^{\rm id} \lesssim 0.8$ or $\Delta
N_{\text{eff}} \lesssim 0.5$, while a mass $m_s = 5$~eV requires
$T_s/T_\nu^{\rm id} \lesssim 0.6$ or $\Delta N_{\text{eff}}
\lesssim 0.2$. 

Our bounds can hopefully be used for constraining particle-physics-motivated models
with three active and one sterile neutrinos, as those investigated
recently in order to explain possible anomalies in neutrino
oscillation data. Many of these models can be immediately localized in
our figures \ref{figB} or \ref{figC}. For sterile neutrinos or other particles
which do not fall in the thermal or DW category, a good approximation consists
in computing
their velocity dispersion and localizing the model in our figure \ref{figD}
\footnote{However, this approximation could be not so good when the
  distribution $p^2 f(p)$ of the non-thermal relic peaks near $p=0$,
  as if part of these relics were actually cold, see
  \cite{Boyarsky:2008prep}.}.  Future neutrino oscillation experiments
are expected to test the self-consistency of the standard
three-neutrino scenario with increasing accuracy. If anomalies and
indications for sterile neutrinos tend to persist, it will be
particularly useful to perform joint analysis of oscillation and
cosmological data, using the lines of this work for the latter part.

\section*{Acknowledgments}
\nopagebreak

M.A.A. would like to thank the International Doctorate on
AstroParticle Physics (IDAPP) for financial support, as well as
LAPTH for hospitality during the realization of most of this work.
M.A.A. and J.L. thank INFN for supporting a visit to the Galileo Galilei
Institute for Theoretical Physics, in which this project was
initiated.
J.L. also acknowledges the support of the EU 6th
Framework Marie Curie Research and Training network ``UniverseNet''
(MRTN-CT-2006-035863). Numerical simulations were performed on the
MUST cluster at LAPP, Annecy (IN2P3/CNRS and Universit\'e de Savoie).



\raggedright
\bibliographystyle{phaip}

\end{document}